%% file: n-resonance.tex
\documentclass{article}
\usepackage{graphicx} 
\usepackage{amsmath} 
\usepackage{amssymb} 
\usepackage{color}

\newcommand{\bbibitem}{\bibitem}

\newcommand{\llabel}[1]{{\label{#1}}}

\newcommand{\ffoot}[1]{}
\newcommand{\hhh}{}

\renewcommand{\r}[1]{(\ref{#1})}
\font\tenmsb=msbm10
\font\sevenmsb=msbm7
\font\fivemsb=msbm5
\newfam\msbfam
\textfont\msbfam=\tenmsb
\scriptfont\msbfam=\sevenmsb
\scriptscriptfont\msbfam=\fivemsb
\def\Bbb#1{{\fam\msbfam\relax#1}}
\newcommand{\bi}{\begin{itemize}}
\newcommand{\ei}{\end{itemize}}
\newcommand{\bd}{\begin{description}}
\newcommand{\ed}{\end{description}}
\newcommand{\be}{\begin{enumerate}}
\newcommand{\ee}{\end{enumerate}}

\renewcommand{\i}{\item}

\newcommand{\bqn}{\begin{eqnarray}}
\newcommand{\eqn}{\end{eqnarray}}
\newcommand{\eqnn}{\nonumber\end{eqnarray}}
\newcommand{\eqnl}[1]{\llabel{#1}\end{eqnarray}}

\newcommand{\nn}{\nonumber}
\newcommand{\noi}{\noindent}
\newcommand{\ba}[1]{\begin{array}{#1}}
\newcommand{\ea}{\end{array}}

\newcommand{\R}{\Bbb{R}}
\newcommand{\C}{\Bbb{C}}
\newcommand{\N}{\Bbb{N}}

\newcommand{\fine}{\end{document}}

\def \trait (#1) (#2) (#3){\vrule width #1pt height #2pt depth #3pt}
\def \qed{\hfill
        \trait (0.1) (6) (0)
        \trait (6) (0.1) (0)
        \kern-6pt   
        \trait (6) (6) (-5.9)
        \trait (0.1) (6) (0)
\medskip}
\def \qedmio{\hfill
             \trait (8) (8) (-0.1)
             \medskip}
\def \quadp{{\Huge $\qedmio$}}
\def \quadv{{\Huge $\qed$}}

\newtheorem{ml}{\bf Lemma}
\newtheorem{Theorem}{\bf Theorem}

\newtheorem{mrem}{\bf \underline{{\sl Remark}}}
\newtheorem{mcc}{\bf Corollary}
\newtheorem{Definition}{\bf Definition}
\newtheorem{mpr}{\bf Proposition}

\newcommand{\bt}{\begin{Theorem}}
\newcommand{\et}{\end{Theorem}}
\newcommand{\bl}{\begin{ml}}
\newcommand{\el}{\end{ml}}
\newcommand{\bp}{\begin{mpr}}
\newcommand{\ep}{\end{mpr}}
\newcommand{\bc}{\begin{mcc}}

\newcommand{\ec}{\end{mcc}}
\newcommand{\bdeff}{\begin{Definition}}
\newcommand{\edeff}{\end{Definition}}
\newcommand{\brem}{\begin{mrem}\rm}
\newcommand{\erem}{\end{mrem}}

\newcommand{\proof}{{\bf Proof. }}

\newcommand{\al}{\alpha}

\newcommand{\ph}{\phi}

\newcommand{\wpsi}{\widetilde{\psi}}
\newcommand{\wH}{\widetilde{H}}

\newcommand{\F}{{\cal F}}

\newcommand{\J}{{\cal J}}

\renewcommand{\H}{{\cal H}}

\newcommand{\sceq}{Schr\"{o}dinger equation\ }  
\newcommand{\sceqn}{Schr\"{o}dinger equation}  
\newcommand{\neigh}{neighborhood }

\newcommand{\e}{\mbox{e}} 
\newcommand{\ti}{\mbox{\tiny i}}

\renewcommand{\phi}{\varphi}

\newcommand{\parti}[1]{\frac{\partial}{\partial #1}}

\newcommand{\SSS}{{\cal S}\!\!\!{\cal S}}
\newcommand{\HHH}{{\cal H}\!\!\!\!{\cal H}}
\newcommand{\UUU}{{\cal U}\!\!\!\!{\cal U}}

\begin{document} 

\begin{center} \noindent
{\LARGE{\sl{\bf Resonance of Minimizers for
$n$-level Quantum Systems with an Arbitrary Cost}}}

\vskip 1cm
Ugo Boscain,
\footnote{
This research
has been supported by a Marie Curie Fellowship.
Program  ``Improving Human Research Potential and the 
Socio-economic
Knowledge Base'', Contract Number: HPMF-CT-2001-01479.
}\\
{\footnotesize D\'epartement de
Math\'ematiques, Analyse Appliqu\'ee et Optimisation,\\
Universit\'e de Bourgogne,
9, Avenue Alain Savary B.P. 47870-21078 Dijon Cedex, France}\\

Gr\'egoire Charlot\footnote{Supported by european 
networks
NCN (ERBFMRXCT97-0137) and NACO2 (N.HPRN-CT-1999-00046)} \\
{\footnotesize SISSA-ISAS, via Beirut 2-4, 34014 Trieste 
(Italy)}\\
e-mails {\tt boscain@sissa.it, charlot@sissa.it}

\end{center}

\vspace{.5cm} \noindent \rm

\begin{quotation}
\noindent  {\bf Abstract}
We consider an optimal control problem describing a 
laser-induced
population transfer on a $n$-level quantum system.

For a convex cost depending only on the moduli
of controls (i.e. the lasers intensities),
we prove that \underline{there always exists a minimizer in
resonance}. This permits to justify
some strategies used in experimental physics. It is also quite 
important
because it permits to reduce remarkably
the complexity of the problem (and extend some of our previous 
results
for $n=2$ and $n=3$): instead of looking for minimizers on the
sphere $S^{2n-1}\subset\C^n$ one is reduced to look just for
minimizers on the sphere $S^{n-1}\subset \R^n$.

{\hhh Moreover, for the reduced problem, 
we investigate on the question of existence of strict 
abnormal 
minimizer.} 
\end{quotation}



\section{Introduction}
\ffoot{ricordarsi di dire:
\bi
\i che parliamo della sincronia dei lasers
\i che controcorrente trattiamo il problema downstairs
\ei
}
The problem of designing an efficient transfer of population between 
different 
atomic or molecular levels is crucial in many atomic-physics 
projects \cite{car1,car2,car3,shorebook,bts}.  
Often excitation or ionization 
is 
accomplished by using 
a sequence of laser pulses to drive transitions from each state to the next 
state.  The transfer should be as efficient as possible  in order to 
minimize the effects of relaxation or decoherence that are always present.
In the recent past years, people started to approach the design of laser 
pulses by using Geometric Control Techniques (see for instance
\cite{altafini,daless,rabitz,brokko,khaneja,rama}).\ffoot{magari 
aggiornare ancora un po'}
Finite dimensional closed quantum systems are in fact left 
invariant control systems on $SU(n)$, or on the corresponding Hilbert 
sphere $S^{2n-1}\subset\C^n$, where $n$ is the number of atomic or 
molecular  levels. For these kinds of systems very powerful techniques 
were 
developed  both for what concerns controllability 
\cite{G-a,G-gb,G-jk,G-js,yuri} 
and 
optimal control \cite{agra-book,jurd-book}.\ffoot{citare anche altro}

The most important and powerful tool for the study of optimal trajectories 
is the well known Pontryagin Maximum Principle (in the following PMP, see 
for instance 
\cite{agra-book,jurd-book,pontlibro}). It is a first order 
necessary condition for optimality and generalizes the Weierstra\ss \ 
conditions of Calculus of Variations to problems with non-holonomic 
constraints. For each optimal trajectory, the PMP 
provides a lift to the cotangent bundle that is a solution to a suitable
pseudo--Hamiltonian system.

Anyway, giving a complete solution to an  optimization problem (that for 
us 
means to give an 
\underline{optimal} \underline{synthesis}, see for instance 
\cite{boltianski,libro,brun1,brun2,piccoli-sussmann}) remains 
 extremely difficult for several reasons. First one is faced with the 
problem of integrating a Hamiltonian system (that generically is 
not  integrable excepted for very special costs). Second one should manage 
with ``non Hamiltonian solutions'' of the PMP, the so called 
\underline{abnormal 
extremals}. Finally,  even if one is able to find all the solutions of the 
PMP it remains the problem of \underline{selecting} among them the
\underline{optimal trajectories}.
For these reasons, usually, one can hope to find  a complete solution of 
an optimal control problem  in low dimension only.

This paper is the continuation of a series of papers on optimal control of 
finite dimensional quantum systems \cite{q1,q2}.
The main purpose is to show that for a certain class of  
quantum systems (that contains systems  useful for applications) one 
can reduce remarkably
the complexity of the problem. More precisely we prove that 
for a convex cost depending only on the moduli of controls 
(e.g. {\hhh amplitude} of the lasers):
\bi 
\i there  always exists a minimizer in resonance that  connects a source 
and a target 
defined by conditions on the moduli of the components of the wave 
function (e.g. two eigenstates, see Theorem \ref{t-main-q1}). 
Roughly speaking to be in resonance means to use lasers
 oscillating with a frequency equal to the difference of energy between  
the levels  that the  laser is coupling.
As a 
consequence one gets a reduction of the dimension of 
the problem from ${2n-1}$ to $n-1$ ($n$ being 
the number of energy levels), see Corollary \ref{corollary-dim}. From 
a physical point of view this means that one is reduced to look just for 
the amplitudes of the lasers; 

\i {\hhh for the reduced system in dimension $n-1$, 
we prove that close to any time t of the domain of a given  minimizer, 
there exists an interval of time where the minimizer is not strictly 
abnormal (see Theorem 
\ref{t-main-q3}). 
This result is a first step in trying to prove the \underline{conjecture} 
that \underline{for the reduced problem there are no} \underline{strictly
abnormal minimizer.}}
\ei
This extends some of our previous results (see \cite{q1,q2}).

Here we are considering a class of systems on which it is possible to 
eliminate the so called drift term. 
This  includes $n$-level 
quantum systems in the rotating wave function
approximation (RWA) and in which each laser couples only close levels. 
For 
this kind of systems  our reduction is crucial to give a complete 
solution for systems with $n=3$ for several costs 
interesting for applications. Moreover it gives some hope to find the 
time optimal synthesis for systems with four levels and bounded lasers. 
Finally it  is of great help in finding numerical solutions to problems 
with 
$n\geq4$. \ffoot{vedere se parlare della Q2, vedere se parlare di quello 
che si 
puo' dire sulla parte reale dei controlli e di K+P}

\medskip

The paper is organized as follows.
In Section \ref{s-model} we introduce the physical model. In order to 
characterize controllability, in Section 
\ref{s-graphs} we associate a topological graph to the system in 
a unique way. 
In Section \ref{s-problem} we formulate the 
optimal 
control problem, while in Section \ref{ss-cost} we discuss the costs that 
are 
interesting for applications. In Section \ref{s-questions} we introduce 
some key definitions and we state our main 
questions.

Section \ref{edt} is devoted to  recall how to eliminate the drift 
term from the 
control
system, by means of a unitary time-dependent change of coordinates plus a
unitary change of controls (interaction picture). This permits to write 
the system in ``distributional'' form.

Our main results are stated in Section \ref{s-main-results} (see Theorems 
\ref{t-main-q1}, 
\ref{t-main-q3}). 
{\hhh Beside stating that there are always minimizers in resonance and 
studying 
strictly abnormal minimizers
for the reduced 
problem,} we 
investigate also the 
question if every minimizer is in resonance (see Theorem 
\ref{t-main-q2}). More precisely, we state that, under the assumption
that the cost function is strictly increasing with respect to the moduli 
of the controls, every minimizer is \underline{weakly-resonant} in a 
suitable 
sense.

In Section \ref{Q12} we prove the main results about resonance. To 
investigate resonance, the key point is to identify the components 
of the controls responsible of the evolution of the moduli of the 
components of the wave function. The difficulty comes from the fact that, 
when a coordinate is zero, the dimension of the ``distribution'' may 
fall. This obliges to divide the domain of a minimizer in suitable 
sub-domains (see Section \ref{s-intervals}). In the first part of Section 
\ref{Q12} we prove our main results, while in Section 
\ref{s-transversality} we give an alternative geometrical interpretation. 
In Section \ref{s-pure-states} we investigate the question if it is 
possible to join every couple of points in $S^{2n-1}\subset\C^n$  by a 
resonant minimizer. In Section 
\ref{examples} we give an example of a minimizer (for a non strictly 
increasing cost) which is not 
weakly-resonant and we propose an open question.

{\hhh The most technical part is  Section \ref{Q3} where we 
investigate about strictly abnormal minimizers
for the 
reduced problem on the sphere 
$S^{n-1}\subset \R^n$. 
For trajectories whose coordinates are never zero, our results are just a 
consequence of the fact that the dimension of the distribution is the 
same as the dimension of the state space. But again the difficulty is 
coming from the fall of the dimension when some 
coordinates are zero.}

\subsection{The Model}
\llabel{s-model}
In this paper we consider closed finite dimensional quantum systems 
in the rotating wave function
approximation (RWA). More precisely  we 
consider systems whose
dynamics is  governed by the time 
dependent Schr\"odinger equation (in
a system of units such that $\hbar=1$):    
\bqn
\left\{
\ba{l}
i\mbox{{\Large $\frac{d\psi(t)}{dt}$}}=\H(t)\psi(t):=(D+V(t))\psi\\\\
\psi(.):=(\psi_1(.),...,\psi_n(.)):\R\to\C^n,~~~
\sum_i|\psi_i|^2=1,
\ea\right.
\eqnl{se}
where it holds:
\bd
\i{(H1)}
$D=diag(E_1,...,E_n)$ and $V(t)$ is an Hermitian matrix
$(V(t)_{j,k}=V(t)_{k,j}^\ast)$, measurable as function of $t$,  whose 
elements $V(t)_{j,k}$ are 
\underline{either identically zero}
or \underline{controls}.
\ed
Here $(^\ast)$ indicates the complex conjugation involution.
$E_1,...,E_n$ are the energy levels of the quantum system and 
the controls $V_{j,k}(.)$, that we assume  to be 
$\C$-valued measurable functions, are different
from
zero only in a fixed interval
$[0,T]$. They are
connected to the physical parameters by $V_{j,k}(t)=\mu_{j,k}{{\cal 
F}}_{j,k}(t)/2$, $j,k=1,...,n$, with
${\cal F}_{j,k}$ the external pulsed fields (the lasers) and
$\mu_{j,k}=\mu_{k,j}>0$ the couplings
(intrinsic
to the quantum system). In the following we say that \underline{two levels 
$E_j$,$E_k$  
are coupled if $V_{j,k}$ is a control} (and not zero).
Moreover if all the couplings $\mu_{j,k}$
are equal to a constant $\mu$ (that we normalize to $1$) we say that the 
system is \underline{isotropic}. Otherwise we say that the system is 
\underline{non-isotropic}.
\brem
The term $D=diag(E_1,...,E_n)$ in equation \r{se} is called
\underline{drift} and it will be eliminated in Section 
\ref{edt}, thanks to the fact that we are in the rotating wave 
approximation and that
each control couples only two levels.
\erem
\brem
This finite-dimensional problem can be thought as the reduction of an
infinite-dimensional problem in the following way.
We start with a Hamiltonian which is the sum of a ``drift-term'' $D$,
plus a time dependent potential $V(t)$ (the control term, i.e. the
lasers). The drift term is assumed to be diagonal, with eigenvalues
(energy levels)
$...>E_3>E_2>E_1.$
Then in this spectral resolution of
$D$, we assume the control term $V(t)$  to couple only a finite number of 
energy levels by pairs. Let $\{E_j\}_{j\in I}$ ($I$ finite) be the set of 
coupled 
levels. The projected problem in the
eigenspaces corresponding
to $\{E_j\}_{j\in I}$ is completely decoupled and it is described by 
\r{se}, (H1).
\erem
The function $\psi:[0,T]\to S^{2n-1}\subset \C^n$ solution of the \sceq 
\r{se} is called the \underline{wave function}. The physical meaning of 
its components $\psi_j(t)$ is the following. For time $t<0$ and 
$t> T$ (where the controls are zero),  $|\psi_j(t)|^2$  is the
probability of measuring energy $E_j$. 
Notice that, in the intervals of time where $V$ is 
identically 
zero, we have:
$$
\frac d{dt}|\psi_j(t)|^2=0,~~~j=1,...n.
$$
In the following a state for which we have $|\psi_j(t)|=1$ for some $j$ 
will be called an \underline{eigenstate}.
\brem 
This model is physically reasonable in the case in which:
\bi 
\i the number of energy levels is not too big, and they 
are distinct by pairs;
\i there are not too many couplings between the energy levels.
\ei
The most interesting case is perhaps the one in 
which only close levels are coupled.
In this case  $V$ has non null coefficients $V_{j,k}$ only on the
second diagonals (for $j$
and $k$ such that $|j-k|=1$) and the Hamiltonian reads:
\bqn
\H(t)&=&
\left(
\begin{array}{ccccc}
E_{1} & V_{1,2}(t) & 0 & \cdots  & 0 \\
V_{1,2}^{\ast}(t) & E_{2} & V_{2,3}(t) & \ddots  & \vdots  \\
0 & V_{2,3}^{\ast}(t) & \ddots  & \ddots  & 0 \\
\vdots  & \ddots  & \ddots  & E_{n-1} & V_{n-1,n}(t) \\
0 & \cdots  & 0 & V_{n-1,n}^{\ast}(t) & E_{n}
\end{array}
\right). \label{mie}
\eqnl{ham-1}
 In the rest of the paper, we will refer to 
this model as to the \underline{most important example}. 
\erem
\subsection{Controllability and Representation of the System with a 
Graph} 
\llabel{s-graphs}
To each system of the form \r{se}, (H1) one can associate a 
\underline{topological graph} (i.e. a set of points and a set of edges 
connecting the points) in a 
very natural way. The points are associated to the energy levels $E_j$ 
$(j=1,...,n)$ and two points $E_j$, $E_k$ are connected by an edge iff the 
element $V_{j,k}$ is a control. This makes sense also if all the energy 
levels are zero (it happens after elimination of the drift, see Section 
\ref{edt}).

\ffoot{{\bf In 
Figure ..... some example of systems 
and corresponding graphs are portrayed. This representation 
is useful to give conditions of controllability and later one is used to 
classified some kind of completely solvable systems.}}
\ffoot{magari dire 
qualcosa di piu'}

Lifting the problem for the operator of temporal evolution (i.e. on 
$U(n)$) and with 
standard arguments of controllability on compact Lie 
groups and corresponding homogeneous spaces, one
gets the following (see the recent survey \cite{yuri} or the papers 
\cite{G-a,q1,q2,boot,G-gb,G-jk,G-js}):
\bp
The control system \r{se}, under the assumption (H1) is completely 
controllable from any initial to any final condition if and only if the 
corresponding graph is connected.
\ep
In the following we deal with optimal control problems and we assume 
existence of minimizers for every couple of points. Hence we assume:
\bd
\i{(H2)} The graph associated with the control system \r{se},(H1) is 
connected.
\ed
If (H2) does not hold, then all the results of the paper are true  for the 
restricted systems associated to the connected parts of the 
Graph.\ffoot{ricordarsi di trattare la parte K+P coi graphi}
\brem
Notice that to guarantee controllability it is \underline{necessary} 
(but not sufficient) to 
have at least $n-1$ controls $V_{j,k}$ (with $j<k$).
\llabel{r-necessary}
\erem
\subsection{The Optimal Control Problem}
\llabel{s-problem}
In this paper, we are faced with the problem of finding optimal 
trajectories 
for a \underline{convex cost
depending only on} \underline{the moduli of controls}
between a source and a target  defined by conditions on the 
moduli of the components of the wave function. 
These are the most common situations in physics since the squares of 
moduli of the 
components of the wave function represent the probabilities of measures.
More precisely our problem is 
the following:\\\\
{\bf Problem (P)} {\it 
Consider the control system \r{se},(H1),(H2) and 
assume that for time $t=0$ the state of the system
is described by a wave function $\psi(0)$ whose components satisfy
$(|\psi_1(0)|^2,...,|\psi_n(0)|^2)\in S_{in}$, where $S_{in}$ is a
subset of the set:
\bqn
\SSS:=\{(a_1,...,a_n)\in(\R^+)^n: \sum a_i=1\}.
\eqnl{sss} 
We want to determine suitable
controls $V_{j,k}(.),$ $j,k=1,...,n$, defined on an interval 
$[0,T]$, such that for time $t=T$,
the system is described by  the wave function $\psi(T)$ satisfying
$(|\psi_1(T)|^2,...,|\psi_n(T)|^2)\in S_{fin}$, where 
$S_{fin}\subset\SSS $,  
requiring that these controls minimize a
\underline{cost that depends only on the moduli of controls}:
\bqn
\int_{0}^{T}f^0(V(t))~dt.
\llabel{cost-general}
\eqn
}
\brem
Typical sources and targets are eigenstates. For instance if the 
source and the target are respectively the eigenstates $1$ and $n$, 
we have $S_{in}=(1,0,...,0)$,  $S_{fin}=(0,0,...,1)$.
\erem
In the following to guarantee  the  existence of minimizers we assume:
\bd
\i{(H3)} The function $f^0$ (that depends only on the moduli of controls) 
is 
convex. $S_{in}$ and $S_{fin}$ 
are closed subset of $\SSS$. Moreover the 
 moduli of controls are 
bounded: there exist constants $M_{j,k}\geq0$ such that 
$|V_{j,k}(t)|\leq  
M_{j,k},$ for every $t\in[0,T]$.
\ffoot{in realta' 
\bi
\i ci basterebbe 
essentially bounded
\i io metterei anche $f^0>0$ e poi direi quialcosa sul fatto che il tempo 
deve essere fisso se .......tipo energia
\ei
} 
\ed
\brem
In the problem  {\bf (P)}, the final time can be fixed or free depending 
on the problem and on the explicit form of $f^0$ (see below).
\erem

\brem
\llabel{r-petrassiemorto}
Notice that if $f^0$ is convex and depends only on the moduli of controls 
then it is an increasing function (as function of the moduli of 
controls).
Hypotheses (H3) could be relaxed, in particular if we do not require 
existence of minimizers for each couple of points, it is not always 
necessary  to assume that controls 
are bounded or that $f^0$ is convex. Moreover the hypotheses of boundedness 
of 
controls could be changed with suitable hypotheses of growing 
of $f^0$ at infinity.
Anyway these investigations are not the purpose of the paper. The costs on 
which we are interested are those described in the next Section, that are 
convex (some of them strictly). Moreover the corresponding
minimization problems  are always 
equivalent to minimization problems with bounded controls.
\erem
In the following if $V(.)$ is a function satisfying (H1), (H2), 
(H3), and $\psi(.)$ the corresponding absolutely continuous trajectory we 
say that the couple 
$(\psi(.),V(.))$ is an \underline{admissible pair}.
\subsection{The interesting costs}
\llabel{ss-cost}
In this paper, we consider only costs that depend on the moduli of
controls, since these are the  interesting costs from a 
physical point of view, like those described in the following. \\\\
{\bf Energy} 
\bqn
\int_0^Tf^0~dt,~~~f^0=
\sum_{j\leq k} \frac{1}{\mu_{j,k}^2} |V_{j,k}|^2.
\eqnl{nonis-en}
This cost is proportional to  the energy of
the laser pulses.
After elimination of the drift (see Section \ref{edt}), the problem {\bf 
(P)}, with this cost 
is a sub-Riemannian problem or a singular-Riemannian problem depending 
on the number of controls (see (H1)) and was studied in
\cite{q1,q2}, in the isotropic 
case ($\mu_{j,k}=\mu$), for $n=2,3$ and in the case in which the 
Hamiltonian is  given
by \r{ham-1}. For sub-Riemannian and singular-Riemannian geometry see for 
instance \cite{bellaiche,gromov,montgomery}.
y
For this cost the final time must be fixed otherwise there are not 
solutions to the minimization problem.
We recall that a minimizer for this cost is parametrized with constant
velocity ($\sum_{j\leq k} \frac{1}{\mu_{j,k}^2} |V_{j,k}|^2=const$)  and 
it is also a minimizer 
for the
sub-Riemannian length:
\bqn
\int_0^Tf^0~dt,~~~f^0=\sqrt{  \sum_{j\leq k}   
\frac{1}{\mu_{j,k}^2} |V_{j,k}|^2}. 
\eqnl{nonis-length}
This cost (\ref{nonis-length}) is invariant by re-parametrization, hence 
the final time
$T$ can be equivalently fixed or free. 
For the costs \r{nonis-en},\r{nonis-length} \underline{the controls are 
assumed 
to be 
unbounded}. Anyway  if the final
time $T$ is fixed in such a way the minimizer is parametrized by
arc-length ($\sum_{j\leq k}  \frac{1}{\mu_{j,k}^2} |V_{j,k}|^2=1$), then 
minimizing the cost 
\r{nonis-en} is
equivalent to minimize time with moduli of controls constrained in the 
closed set: 
\bqn
\sum_{j\leq k} \frac{1}{\mu_{j,k}^2} 
|V_{j,k}|^2\leq1. \label{tiinculo}
\eqn

In the isotropic case  $\mu_{j,k}=\mu=1$ the energy and the length are 
called respectively \underline{isotropic-energy} and  
\underline{isotropic-length}. They read:
\bqn
\int_0^Tf^0~dt,~~~f^0=
\sum_{j\leq k}|V_{j,k}|^2,\llabel{is-en}\\
\int_0^Tf^0~dt,~~~f^0=\sqrt{  \sum_{j\leq k}|V_{j,k}|^2}.
\eqnl{is-length}
and the set (\ref{tiinculo}) becomes a closed ball.
In the non-isotropic case we call \r{nonis-en} and \r{nonis-length} 
respectively  
\underline{non-isotropic-energy} 
and
\underline{non-isotropic-length}.\\\\
{\bf Isotropic and Non-isotropic Area} 
\bqn
\int_0^Tf^0~dt,~~~f^0=
\sum_{j\leq k} \frac{1}{\mu_{j,k}} |V_{j,k}|,~~~~~ 
\mbox{(isotropic case
$\mu_i=\mu=1$)} \eqnl{nonis-area} These costs are proportional to the area
of the laser pulses {\large$\int^T_0$}$\sum_{j\leq
k}|\F_{j,k}(t)|~dt$. They are invariant by re-parametrization.  
Minimizing these costs is equivalent to minimize time with the following
constraint on controls: 
$$ \sum_{j\leq k}\frac{1}{\mu_{j,k}^2}|V_{j,k}|\leq1. $$
{\bf Time with Bounded Controls}\\ 
If we want to minimize time having a bound on the
maximal amplitudes of the lasers, $|{\cal F}_{j,k}|\leq \nu_{j,k}$ 
($j\leq 
k$) then we get the
bounds on $V_{j,k}$: $|V_{j,k}|\leq \mu_{j,k}\nu_{j,k}/2$. With the change 
of
notation $\mu_{j,k}\nu_{j,k}/2\to\mu_{j,k}$ we get:  
\bqn 
|V_{j,k}|\leq \mu_{j,k}.
\eqnl{nonis-bound} 
In this case we have $f^0=1$. 
Now the isotropic case $\mu_{j,k}=\mu=1$ 
 corresponds to an isotropic system in which all the lasers have the 
same bound on the amplitude. 

In this case the problem is equivalent (up to a re-parametrization) to 
minimize: 
\bqn
\int_0^Tf^0~dt,~~~
f^0=\max_{j\leq k}\left|\frac{V_{j,k}}{\mu_{j,k}}\right|.
\eqn
In the following figure the level sets $f^0=1$ for the $n=3$ case for the 
Hamiltonian \r{ham-1} are drawn (of course in $\R^2$ instead of $\C^2$). 
If we consider the equivalent problems of 
minimizing time with constrained  controls, then controls should lie in
the compact regions whose borders are these level sets. 
\begin{center}
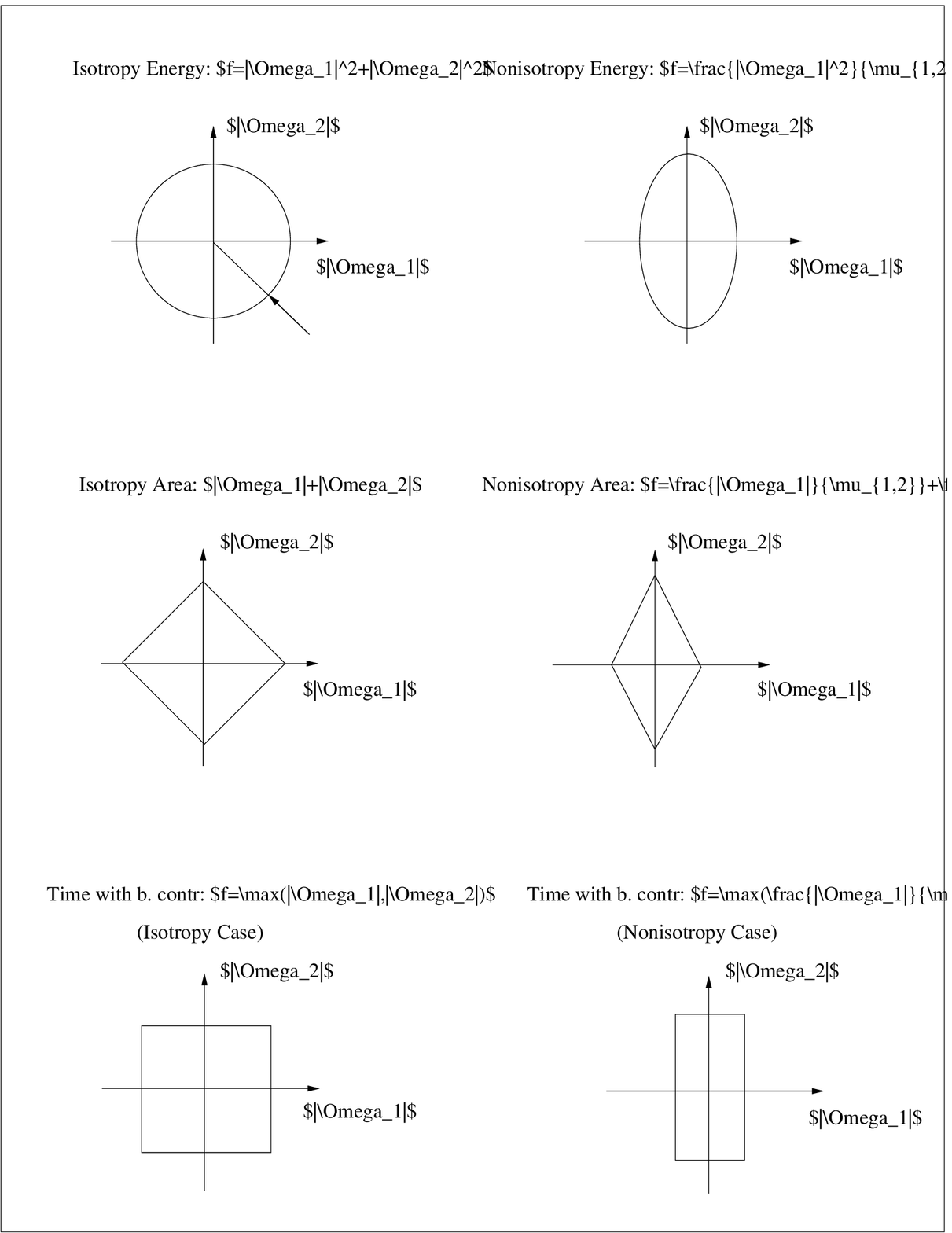
Figure 1: level $f^0=1$ for different cost functions
\end{center} 

\subsection{Main Questions}
\llabel{s-questions}
Although the system \r{se}, (H1) has a lot of symmetries, and good 
properties, the problem {\bf (P)} in general is very difficult to 
solve also for the costs described above, for the reasons explained in the 
introduction:
\bi
\i problem of integrability of the 
 Hamiltonian system associated with the PMP,
\i presence of abnormal extremals,
\i  problem of selecting the
optimal trajectories among those satisfying the PMP.
\ei
\ffoot{The first 
difficulty that 
one meets is the problem of  
integrating the Hamiltonian System associated with the Pontryagin Maximum 
Principle (in general it is not integrable!). We recall that the
Pontryagin Maximum Principle is a first order necessary condition for
optimality and can be seen as a generalization of the classical necessary
conditions
for optimality, of the calculus of variations, to trajectories with
constrained velocities. 
Then, even if one is able to find extremal trajectories, for a very  
special cost, it remains the problem of selecting the 
optimal ones.}
Up to now the problem is solved for $n=2$ (for the costs described in 
Section \ref{ss-cost}, that, in this case, are all equivalent) 
and for $n=3$, for the Hamiltonian given by \r{ham-1} 
for the isotropic energy (see \cite{q1,q2}).
In both cases 
\underline{optimal 
controls appear to be in resonance} (with
the difference between the energy levels  that they are coupling)
and \underline{every strictly} \underline{abnormal extremal is not 
optimal}. In this 
paper we generalize these results for more general systems and costs.
Let us first introduce the definition of resonance and abnormal extremals:
\bdeff {\bf (Resonance)}
Consider the control system \r{se},(H1) and an admissible pair 
$(\psi(.),V(.))$  defined in an
interval $[0,T]$.  We say that the couple  $(\psi(.), V(.))$ is 
in 
\underline{resonance} (or is \underline{resonant}) if the controls 
$V_{j,k}(.)$  have a.e. the form:
\bqn 
&&V_{j,k}(t)=U_{j,k}(t)e^{i[(E_{j}-E_{k})t +\pi/2+\phi_{j,k}]}, 
\mbox{ ~~~~where:}\llabel{res0}\\
&&U_{j,k}(.):[0,T]\to\R, \;\;U_{j,k}=-U_{k,j},\llabel{res1}\\
&&\ph_{j,k}:=arg(\psi_j(0))-arg(\psi_k(0))\in[-\pi,\pi]
\llabel{res2}
\eqn
In formula \r{res2}, if $\psi_j(0)=0$, then $arg(\psi_j(0))$ is intended 
to 
be an arbitrary number. 
\edeff
\brem
From a physical point of view, to be in resonance means to use lasers 
(described by complex functions) oscillating with a frequency given by 
$\omega/(2\pi)$ where $\omega$ is the difference of energy between  the 
levels  that the  laser is coupling. Notice that in formula (\ref{res1}) 
$U_{j,k}(.)$ takes real values (it describes the amplitude of the lasers).

In the definition of resonance,
the phases  $\phi_{j,k}$ sometimes are important and sometimes not, 
depending on the explicit form of the control term $V(t)$ in \r{se} and on 
the 
choice of the source and the target. 
For instance if we are 
considering the problem described by the Hamiltonian \r{ham-1} and we 
start from an eigenstate (e.g. $|\psi_1(0)|=1$) then all the phases 
$\phi_{j,k}$ appearing in formula \r{res0}
are arbitrary. In fact in this case we have $n-1$ arbitrary numbers
$arg(\psi_j(0))$ and $n-1$ controls. Therefore in this case it is 
not necessary  to synchronize the phases of the lasers between them.

On the other side, if we start from a state that is not an eigenstate or 
we have 
more controls than the minimum number necessary to guarantee 
controllability (see Remark \ref{r-necessary}), to be in resonance 
we need to synchronize the lasers according to formula \r{res2}. 

In any case notice that a global 
factor of phase in front of $\psi$ is not important because it can be 
eliminated with a unitary transformation. Hence all the phases 
$\phi_{j,k}$ can be shifted by an arbitrary factor $\al$. 
\erem
\brem
{\hhh To prove that a minimizer corresponds to controls in 
resonance, 
is very important because, as we will see after elimination of the drift,
this permits us to reduce 
the dimension of the problem from the complex sphere $S^{2n-1}\subset\C^n$ 
to the
real sphere $S^{n-1}\subset \R^n$ (See the Corollary  
\ref{corollary-dim} in 
Section \ref{s-main-results}). 
This reduction of dimension is crucial in finding complete solutions to 
the optimal control problem in many cases. For instance for the class of 
problem described by the Hamiltonian 
\r{mie}:
\bd
\item[A.] an isotropic or non-isotropic minimum time problem with
bounded controls for a 3-level
system, is a problem in
dimension
$5$. In this case, since the dimension of the state space is
big, the problem of finding extremals and selecting optimal trajectories
can be extremely hard.
The possibility of proving that one can restrict to minimizers that are in
resonance (and
this is the case!) permits us to reduce the problem to a bi-dimensional
problem, that can be solved with standard techniques (see for instance
\cite{libro,jurd-book,due,suss});
\item[B] a nonisotropic minimum energy problem for a  3-level system is 
naturally lifted  to a left invariant sub-Riemannian problem on the group 
$SU(3)$. This problem cannot be solved with the techniques used in 
\cite{q2} for the isotropic case, because now the cost is built with a 
``deformed Killing form''. Anyway if one can restrict to minimizers that 
are in resonance, the problem is reduced to a contact sub-Riemannian 
problem  on $SO(3)$, that does not have abnormal extremals (since it is 
contact) and the corresponding Hamiltonian system is completely 
integrable (since it leads to a left 
invariant Hamiltonian system on a Lie group of dimension 3).
\ed
Items {\bf A.} and {\bf B.} are studied in a 
forthcoming paper.
\bd
\item[C.] An isotropic or nonisotropic minimum energy problem for 
a  4-level system is a problem on $S^7$ or can be lifted to a left 
invariant sub-Riemannian problem on $SU(4)$. In this case 
we \underline{conjecture} that \underline{the Hamiltonian system} 
\underline{given by 
the 
PMP} \underline{is not Liouville integrable} both before and after the 
reduction to real 
variables. Anyway  this reduction can be crucial in looking for numerical 
solution to the PMP (it is of course easier making numerical computation 
on  $S^3$ than on $S^7$!). See 
\cite{sevilla} for some numerical solutions to this problem.
\ed
Moreover proving that there always exists a minimizer in resonance  
permits us to justify some strategies used in experimental physics.} 
\erem
To introduce the concept of abnormal extremals, we have to state the PMP, 
that is a necessary condition for optimality. 
The proof can be found for instance in
\cite{agra-book,jurd-book,pontlibro}.\\\\
\noi{\bf Theorem (Pontryagin Maximum Principle)}
{\it Consider a control system of the form $\dot x=f(x,u)$ with a
cost of the form $\int_0^Tf^0(x,u)~dt$, and initial and final 
conditions given by 
$x(0)\in M_{in}$, $x(T)\in M_{fin}$,
where $x$
belongs to  a manifold $M$ and $u\in U\subset \R^m$. Assume moreover that
$M,~f,~f^0$ are smooth and that $M_{in}$ and $M_{fin}$ are smooth 
submanifolds of $M$.
If the couple $(u(.),x(.)):[0,T]\subset\R\to U\times M$ is optimal, then
there
exists
{\sl a never vanishing field of covectors along} $x(.)$, that is an
absolutely
continuous function $(P(.),p_0):t\in [0,T]\mapsto (P(t),p_0)\in
T^\ast_{x(t)}M\times \R$
(where $p_0\leq0$ is  a constant) such that:
\begin{description}
\item[i)]
$\dot x(t)=$ {\large $\frac{\partial \HHH
}{\partial P}$}$(x(t),P(t),u(t))$,
\item[ii)] $\dot P(t)=-${\large$\frac{\partial \HHH
}{\partial x}$}$(x(t),P(t),u(t))$,
\ed
where by definition:
\bqn
\HHH (x,P,u):=<P, f(x,u)>+p_0f^0(x,u).
\eqnl{HAMHAMHAM}
Moreover:
\bd
\item[iii)] $\HHH (x(t),P(t),u(t))=\HHH_M(x(t),P(t))$, for a.e.
$t\in[0,T]$,\\
where $\HHH_M(x,P):=\max_{v\in U}\HHH(x,P,v).$
\item[iiii)] $\HHH_M(x(t),P(t))=k\geq0$, where $k$ depends on the 
final time (if it is fixed) or $k=0$ if it is free.
\item[v)] $<P(0),T_{x(0)}M_{in}>=<P(T),T_{x(T)}M_{fin}>=0$ (transversality 
conditions).
\end{description}
}
The real-valued map on $T^*M\times U$, defined in \r{HAMHAMHAM} is called
\underline{PMP-Hamiltonian}. A trajectory $x(.)$ (resp. a couple 
$(u(.),x(.))$) satisfying conditions
{\bf i)},
{\bf ii)},
{\bf iii)} and 
{\bf iiii)} is called an \underline{extremal} (resp. \underline{extremal 
pair}.)
If $x(.)$ satisfies {\bf i)},
{\bf ii)},
{\bf iii)} and
{\bf iiii)}
with $p_0=0$ (resp. $p_0<0$), then it is called an 
\underline{abnormal extremal} (resp. a \underline{normal extremal}).
It may happen that an extremal $x(.)$ is both normal and abnormal. So it 
makes sense to speak of \underline{strictly abnormal extremals} (extremals 
that are abnormal but not normal).

\brem
Notice that the definition of abnormal extremal does not depend on the 
cost but only on the dynamics (in fact if $p_0=0$, the cost disappears in 
\r{HAMHAMHAM}).
After elimination of the drift (see Section \ref{edt}) our control system 
\r{se} will be 
transformed into a distributional system, that is a system of the form 
$\dot x=\sum u_j F_j(x)$. For these systems abnormal extremals are also 
\underline{singularities of the end point mapping}, and have very special 
features (see for instance the recent monograph \cite{bonnard-book}).
\erem
\brem
In this paper we are dealing only with non-state dependent costs, i.e. 
$f^0(x,u)=f^0(u)$.
\erem
\ffoot{ The problem {\bf (P)}, with this cost 
is a sub-Riemannian problem (for sub-Riemannian geometry see for 
instance.....) and was studied in \cite{q1,q2} for $n=2,3$.} 
Consider the problem {\bf (P)} under the assumption (H3). Hence a 
minimizer  
exists. The main purpose of this paper is to answer to the following 
questions:
\bd
\i[Q1] Does there 
exist a minimizer that corresponds to controls  in resonance? 
\i[Q2] Once we have given a positive answer to question {\bf Q1}, we 
would like to answer the question: are  all 
the minimizers of the problem {\bf (P)}  in resonance?
\i[Q3] {\hhh Do there exist strictly abnormal minimizers?}
\ed
{\hhh The answers to questions {\bf Q1} is yes, while \underline{{\bf
Q3} is still an open question}, although we are 
able to prove a result that strongly restricts the set of candidates 
strictly abnormal minimizers.}

These results are formalized in Theorems in  Section \ref{s-main-results}.

The answer to question {\bf Q2} is in general no, but for an increasing 
cost we show that every minimizer is in \underline{weak-resonance} (again 
this will be formalized in a Theorem after elimination of the drift). To define 
this notion, in the next Section we introduce some notations that will be 
also useful in 
Section \ref{Q12} to answer {\bf Q1} and {\bf Q2}.\ffoot{MA ABBIAMO 
DEFINITO CHE CAZZO VUOL DIRE STRICTLY 
INCREASIND?????????? vedi remark \ref{r-petrassiemorto}} 

Finally another interesting question for physicists is:
\bd
\i[Q1']
is it possible to join two arbitrary states 
$\psi^1$ and $\psi^2$ by an optimal trajectory in resonance?
\ed
The answer is, of course, in 
general, no. But there are
couples of points for which the answer is yes. In particular if we
consider eigenstates, it is true (see Section \ref{s-pure-states}).

\subsubsection{Weak-Resonance}
\llabel{s-intervals}
To introduce this concept, we need some notations.

Consider an admissible pair $(\psi(.),V(.))$ and define the following 
subset of $]0,T[$:
\bqn
Bad_{j,k}=\{t\in]0.T[,\mbox{ s.t. }\psi_j(t)=0\mbox{ or }\psi_k(t)=0\}.
\eqnl{bad1}
Since $\psi(.)$ is continuous, the sets 
$]0,T[\setminus Bad_{j,k}$ are open. Then they can be expressed as a 
countable union of \underline{open maximal intervals}:
\bqn
]0,T[\setminus Bad_{j,k}=:\bigcup_l I_{j,k,l},~~~j,k=1,...,n,~~~ 
l=1,...m,\mbox{ where } \underline{m\in\{0\}\cup{\N}\cup\{\infty\}}.
\eqnl{bad2}
In other words $I_{j,k,l}$ are the maximal open intervals on which 
$\psi_j(t)\psi_k(t)\neq0$.
\brem
In general $\overline{ \cup_l I_{j,k,l} }$ is smaller than $[0,T]$ since 
it may happen that $\psi_j(t)\psi_k(t)=0$ on some interval of positive 
measure. \ffoot{controllare se e' ben detto}
\erem

Now we are ready to give the following:
\bdeff {\bf (weak-resonance)} Consider the control system \r{se},(H1) and 
an admissible pair
$(\psi(.),V(.))$  defined in an
interval $[0,T]$.  We say that the couple  $(\psi(.), V(.))$ is
in
\underline{weak-resonance} (or is weakly-resonant) if in each interval 
$I_{j,k,l}$ (defined 
above) the controls $V_{j,k}(.)$ satisfy a.e.:
\bqn
&&\left.V_{j,k}\right|_{I_{j,k,l}}(t)=U_{j,k,l}(t)e^{i[(E_{j}-E_{k})t 
+\pi/2+\phi_{j,k,l}]},
\mbox{ ~~~~where:}\llabel{wres0}\\
&&U_{j,k,l}(.):[0,T]\to\R,~~U_{j,k,l}=-U_{k,j,l}    \llabel{wres1}\\
&&\phi_{j,k,l}:=arg(\psi_j(a_{j,k,l}))-arg(\psi_k(a_{j,k,l})) \in 
[-\pi,\pi]\llabel{wres2}
\eqn
where $I_{j,k,l}=:]a_{j,k,l},b_{j,k,l}[$. In formula (\ref{wres2}), if 
$\psi_j(a_{j,k,l})=0$, then $arg((\psi_j(a_{j,k,l})))$ is intended to be 
an arbitrary number.
\edeff
\brem
Roughly speaking a control $V_{j,k}(.)$ is weakly-resonant
if it is resonant in each interval of time in which the states that 
it is 
coupling (i.e. $\psi_j$ and  $\psi_k$) are different from zero.

If the cost is not strictly increasing 
(for instance minimum time
with bounded controls) in general there are minimizer that are not in
weak-resonance, see Section \ref{examples} for 
an example and another open question. Finally notice that \underline{a 
resonant minimizer is also  
weakly-resonant}.\ffoot{l'altra open question la lasciamo nell'esempio per 
ora}
\erem
\section{Elimination of the Drift Term}
\llabel{edt}
The great advantage of the  model presented in the previous Section 
(in which each control is complex and couples only two levels) is
that we can eliminate 
the drift term $D=diag(E_1,...,E_n)$ from equation \r{se} (hence getting 
a system in distributional form 
 $\dot x=\sum_j u_j F_j(x)$), 
simply by using the 
\underline{interaction picture}. 
This is made by a unitary
change of coordinates
and a unitary change of controls. 
Since the transformation is unitary, 
$S_{in}$, $S_{fin}$ and the moduli of the components of the wave function 
are invariant. As a consequence the original and the transformed
systems describe exactly the same population distribution. 

Assume that $\psi(t)$ satisfies the \sceq (\ref{se}). Let $\UUU(t)$ be a
unitary time dependent matrix and set $\psi(t)=\UUU(t)\psi'(t)$. 
Then $\psi'(t)$ satisfies the \sceqn: 
\bqn
i\frac{d\psi'(t)}{dt}=\H'(t)\psi'(t), 
\eqnn 
with the new Hamiltonian: 
\bqn
\H'=\UUU^{-1}\H \UUU-i\UUU^{-1}\frac{d\UUU}{dt}.
\eqnl{HUHU} 
If we choose: 
\bqn
\UUU(t)=e^{-iDt},
\eqnl{trasf}
and we recall that $\H=D+V(t)$, we get $\H'=e^{iDt}V(t)e^{-iDt}$,
that is a Hamiltonian whose elements are either zero or can 
be redefined
to be controls. Hence the drift is eliminated.
Finally dropping the primes and including the $i$ in the new Hamiltonian 
 ($\psi'\to\psi$, $H:=-i\H'$)  the \sceq reads:
\bqn
\frac{d\psi(t)}{dt}=H(t)\psi(t).
\eqnl{se-no-i}
Here $H$ is \underline{skew-Hermitian} and its elements are either zero 
or controls. There is no more drift and the dynamics is in  
``distributional'' form, i.e.
$\dot{\psi}_j(t)=\sum_{k=1}^{n}H_{j,k}(t)\psi_k(t)$.

The relation between the old controls $V(t)_{j,k}$ (for equation \r{se}) 
and the new controls $H(t)_{j,k}$ for equation \r{se-no-i} is the
following:
\bqn
V(t)_{j,k}=\left(i e^{-i Dt} H(t)e^{iDt} \right)_{j,k}=
H(t)_{j,k}~ e^{i[(E_k-E_j)t+\pi/2]}.
\eqn
\brem
Notice that the transformation \r{trasf} kills also a 
not-only-diagonal drift. While to kill a time dependent drift we need the
transformation (see \cite{q1}):
$$
\UUU(t)=e^{-i\int_{t_0}^t D(t)~dt}.
$$
\erem

\section{Statement of the Main Results}
\llabel{s-main-results}
With the transformation
described above (formula \r{trasf}, and following), the problem 
{\bf (P)} becomes the problem {\bf (P')} below and the 
answers to 
questions 
{\bf Q1},
{\bf Q2},
{\bf Q3} are contained in the next Theorems, that we are going 
to prove  in Sections \ref{Q12},  \ref{Q3}.\\\\
{\bf Problem (P')} {\it Consider the minimization problem:
\bqn
&&\frac{d\psi(t)}{dt}=H(t)\psi(t)\llabel{mm1}\\
&&\min\int_{0}^{T}f^0\left(H\right)~dt.\llabel{mm2}\\
&&(|\psi_1(0)|^2,...,|\psi_n(0)|^2)\in S_{in}\llabel{mm3}\\
&&(|\psi_1(T)|^2,...,|\psi_n(T)|^2)\in S_{fin}\llabel{mm4}
\eqn
where:
\bi
\i $f^0$  depends only on the moduli of controls,
\i there hold (H1'), (H2), (H3). where  (H1') is the following condition:
\bd
\i{(H1')} the matrix $H(t)$ is  \underline{skew-Hermitian}, 
measurable as 
function of $t$ and its elements are either identically zero or controls.
\ed
\ei
}
\noindent 
We have the following:
\bt {\bf (answer yes to Q1: existence of resonant minimizers)}
Let $(\psi(.),H(.))$, defined in $[0,T]$, be a solution to the
minimization problem {\bf (P')}. Then
there exists another solution $(\bar \psi(.),\bar H(.))$, to the
same minimization problem (with the same source, target and 
initial condition $\psi(0)=\bar\psi(0)$)
that is in resonance:
\bqn
&&\bar H_{j,k}(t)= U_{j,k}(t)e^{i\phi_{j,k}},\mbox{ 
~~~~where:}\llabel{rr0}\\
&&U_{j,k}(.):[0,T]\to\R,~~U_{j,k}=-U_{k,j}\llabel{rr1}\\
&&\ph_{j,k}:=arg(\psi_j(0))-arg(\psi_k(0))\in[-\pi,\pi]
\llabel{rr2}
\eqn
Moreover the arguments of $\bar \psi_j(.)$ do not 
depend on the time.
In formula \r{rr2}, if $\psi_j(0)=0$, then $arg(\psi_j(0))$ 
is intended to be an arbitrary number.
\llabel{t-main-q1}
\et
\bt {\bf (answer to Q2)} 
Let $(\psi(.),H(.))$, defined in $[0,T]$, be a solution to the
minimization problem {\bf (P')}. Then if the cost is strictly increasing,
$(\psi(.),H(.))$ is weakly-resonant:
\bqn
&&\left. H_{j,k}(t)\right|_{I_{j,k,l}}= 
U_{j,k,l}(t)e^{i\phi_{j,k,l}},\mbox{ 
~~~~where:}\llabel{wrr0}\\
&&U_{j,k,l}(.):I_{j,k,l}\to\R,~~U_{j,k,l}=-U_{k,j,l}\llabel{wrr1}\\
&&\phi_{j,k,l}:=arg(\psi_j(a_{j,k,l}))-arg(\psi_k(a_{j,k,l})) \in
[-\pi,\pi]\llabel{wrr2}
\eqn
where $I_{j,k,l}=]a_{j,k,l},b_{j,k,l}[$ are the intervals defined
in Section \ref{s-intervals}. In formula (\ref{wrr2}), if
$\psi_j(a_{j,k,l})=0$, then $arg((\psi_j(a_{j,k,l})))$ is intended to be
an arbitrary number.
\llabel{t-main-q2}
\et
In the rest of the paper we call an \underline{admissible pair}, a couple 
trajectory-control for an optimal control problem satisfying 
(H1'),(H2),(H3). 
\brem
Notice that after elimination of the drift, to be in resonance 
implies to use controls with constant phases.
\erem
An important consequence of Theorem \ref{t-main-q1} is the following. 
\ffoot{(as a
  consequence of the fact that the source and the target are defined by
  conditions on the moduli)} 
If an admissible pair is 
a solution in resonance to the minimization problem, then there exists also 
one for which $\arg(\psi_j(0))=0$,  $j=1,...,n$, i.e. corresponding to 
\underline{real controls and real initial conditions}. In this case 
equation \r{mm1} restricts to 
reals, so $\psi(t)\in S^{n-1}\subset\R^n$.   
Hence it makes sense to give the following:
\bdeff
We call {\bf (RP')}, the problem {\bf (P')} in which all 
the 
coordinates and controls are real.
\edeff
And we have:
\bc
\llabel{corollary-dim}
If ${(\psi(.),H(.))}$ is a solution to the minimization problem {\bf 
(RP')} (that is with $\psi(t)\in S^{n-1}\in \R^n,~~H(t)\in so(n)$) then it 
is also a 
solution 
to the original problem  {\bf (P')} or equivalently, with the 
transformation described in Section \ref{edt}, of the original problem 
{\bf (P)}.  
\ec
{\hhh The following Theorem 
restricts the set of candidates strictly abnormal minimizers, 
for {\bf (RP')}:}
\bt {\bf (partial answer  to Q3 for  {\bf (RP')})} 
\llabel{t-main-q3}
{\hhh 
Let $(\psi(.),H(.))$ be a couple trajectory-control
that is a minimizer for  {\bf (RP')} and assume that there are no 
constraints on the controls ($H(.)_{j,k}:Dom(\psi)\to \R$). 
Then for every $t\in Dom(\psi)$,
there exists an interval $[t_1,t_2]$ arbitrarily close to $t$ (possibly 
non containing $t$),
on which  $(\psi(.),H(.))$ is not a stricly abnormal extremal.}
\et
For properties of abnormal extremals, we refer the reader to 
\cite{agra-book,bonnard-book,libro,montgomery,mont}. 
\brem
{\hhh In Theorem \ref{t-main-q3} the fact that we assume unconstrained 
controls 
is just a technical hypothesis that permits to simplify statements and 
proofs. Notice that for the costs described in Section 
\ref{ss-cost} controls can always be assumed without constraints.}
\erem
{\hhh As we said above, the non existence of stricly abnormal minimizer 
is still a \underline{conjecture}. After the result given by Theorem 
\ref{t-main-q3} to prove this conjecture one is essentially   
left with the following:\\\\
{\bf Problem:} Let 
$(\psi(.),H(.))$ and 
$(\bar \psi(.),\bar H(.))$ be two non strictly abnormal minimizers defined 
respectively on $[t_a,t_b]$ and $[t_b,t_c]$ with $t_a<t_b<t_c$. Is it 
true 
that the concatenation of the two is not strictly abnormal? }


\section{Resonance and Weak-resonance} \llabel{Q12}
\newcommand{\un}{1\hspace{-0.09cm} \mbox{I}}
In this Section we prove Theorems 
\ref{t-main-q1} and 
\ref{t-main-q2}. The key point is to identify the components of the 
control $H_{j,k}(.)$ that are responsible 
of the evolution of $|\psi_j|$ and of
$\arg(\psi_j)$.        
 The difficulty is that these components are well defined only for the 
times such that  
$\psi_j(t)\neq0$ and 
$\psi_k(t)\neq0$ ($\psi_j$ and $\psi_k$ are the states coupled by 
$H_{j,k}$(.)). So we have to split the problem into the intervals 
$I_{j,k,l}$  
defined in Section \ref{s-intervals}.

Let $(\psi(.),H(.))$ be a solution to the minimization problem and 
restrict the problem to the interval $I_\al$, 
$\al=(j,k,l)\in(1,...n)^2\times(1,...,m)$, where $\psi_j$ and $\psi_k$ 
never vanish. Define in $I_\al$:
\bqn
&&\theta_j^\al(t):=arg(\psi_j^\al(t) ),\nn\\ 
&&\rho_j^\al(t):=\left|\psi_j^\al(t)\right|,\nn\\ 
&&\beta_{j,k}^\al(t):=  \theta_j^\al(t)-  \theta_k^\al(t).  \nn 
\eqn
From the definition of $I_{\al}$ the phases are well defined.
Define the functions  $u_{j,k}^\al(t)$ and $v_{j,k}^\al(t)$ in such a way
that:
\bqn
H_{j,k}^\al(t):=\left.H_{j,k}(t)\right|_{I_\al}=:(u_{j,k}^\al(t)+i 
v_{j,k}^\al(t))e^{i\beta_{j,k}^\al(t)    }.
\llabel{eq-uv}
\eqn                     
We have:
\bqn
\left\{
\ba{l}
u_{j,k}^\al=-u_{k,j}^\al,~~~v_{j,k}^\al=v_{k,j}^\al\\
|H_{j,k}^\al|=|u_{j,k}^\al+iv_{j,k}^\al|.
\ea\right.
\llabel{eq-hacca}
\eqn   
Now in $I_\al$:
$$
\dot{\psi}_j(t)=\sum_k H_{j,k}^\al(t)\psi_k(t)=\sum_k
\big({u}_{j,k}^\al(t)+i {v}_{j,k}^\al(t)\big)
|\psi_k(t)|\e^{i\theta^\al_j(t)}.
$$
From that, with a simple computation one gets:
\bqn
\frac{d}{dt}\left(\rho_j^\al(t)\right)&=&\sum_k {u}_{j,k}^\al(t) 
\rho_k^\al(t),\llabel{eq-mod}\\
\frac{d}{dt}\left(\theta_j^\al(t)\right)&=&\sum_k {v}_{j,k}^\al(t)
\rho_k^\al(t)/\rho_j^\al(t).\llabel{eq-arg}                  
\eqn 
These equations show that, on $I_\al$, 
$u_{j,k}^\al$ and 
$v_{j,k}^\al$ are responsible respectively of the evolution of 
$\left|\psi_j\right|$ and $arg(\psi_j)$.

From the original minimizer $(\psi(.),H(.))$ we now define a new minimizer 
by means of two consecutive transformations:
\bi
\i the first in which we set at zero all the $v^\al$-components of the 
control and we get a new admissible pair connecting $S_{in}$ and $S_{fin}$ 
having a strictly smaller cost for a strictly increasing $f^0$ and that 
is weakly-resonant. This proves Theorem \ref{t-main-q2};
\i the second in which $H_{j,k}$ is set to zero where $\psi_j\psi_k=0$ and 
in which all the phases $\beta_{j,k}^\al$ are set to 
$arg(\psi_j(0))-arg(\psi_k(0))$. This transformation does not increase the 
cost and produce a control in resonance. This proves Theorem 
\ref{t-main-q1}.
\ei
These transformations are realized in the following.\\\\
Let $(\psi(.),H(.)$ be the original minimizer and let $u^\al_{j,k}$ and 
$v^\al_{j,k}$ the corresponding components of the control in $I_\al$, 
defined in formula \r{eq-uv}. 
Define a new admissible pair $(\bar\psi(.),\bar H(.))$ by 
setting in 
$I_\al$:
\bqn
\left\{\ba{l}
\bar u^\al_{j,k}=u^\al_{j,k}\\
\bar v^\al_{j,k}\equiv0
\ea\right.
\eqn
Notice the important point that  $\bar v^\al_{j,k}\equiv0$ implies that 
the corresponding  
$\bar \beta^\al_{j,k}$ are constants (see equation \r{eq-arg}) and they 
will be 
considered arbitrary phases. In other words  $(\bar\psi(.),\bar H(.))$ is 
the admissible pair corresponding to the control:
\bqn
\bar H_{j,k}(t)=\left\{\ba{l}
H_{j,k}(t)\mbox{ if }t\notin  \cup_\al I_\al\\
u^\al_{j,k}(t)e^{i \bar \beta_{j,k}^\al}\mbox{ if }t\in I_\al\mbox{ for 
some $\al$},
\ea\right.
\eqnl{kkk1}
where $ \bar \beta_{j,k}^\al$ are constant arbitrary phases.
The corresponding trajectory $\bar\psi(.)$ in each 
$I_\al$ satisfies:
\bqn
\left\{\ba{l}
\bar\psi_j(t)=\bar\rho_j^\al(t) e^{i\bar \theta^\al_j}\\
\bar\psi_k(t)=\bar\rho_k^\al(t) e^{i\bar \theta^\al_k}\\
\ea\right.
\eqnl{kkk2}
with   $\bar \theta^\al_j$ and  $\bar \theta^\al_k$ constant and 
such that 
$\bar \theta^\al_j-\bar \theta^\al_k=\bar\beta^\al_{j,k}.$
Now since  $\bar H_{j,k}=H_{j,k}$ if $t\notin \cup_\al I_\al $ and 
the $u_{j,k}$ component is always the same for every $\al$, it follows 
that 
$|\bar\psi_j(t)|=|\psi_j(t)|$ for every $t\in[0,T]$. This means that 
$\psi(.)$ 
and $\bar \psi(.)$ connect the same source and target (recall that they 
are 
defined by conditions on moduli). By construction $(\bar\psi(.),\bar 
H(.))$  is in \underline{weak-resonance}.\\\\
{\bf Proof of Theorem \ref{t-main-q2}} By contradiction assume that:
\bi
\i $f^0$ is strictly increasing;
\i $(\psi(.),H(.))$ is not in weak-resonance that means 
$v^\al_{j,k}(t)\neq0$ on some set of positive measure $A\subset[0,T]$ and 
for some 
indexes 
$j,k,\al$. 
\ei
In this case from equation \r{eq-hacca}  we have:
$$
|\bar H_{j,k}(t)|<|H_{j,k}(t)|\mbox{ for each t}\in A,
$$
that means:
$$
\int_0^Tf^0(\bar H(t))dt<\int_0^Tf^0(H(t))dt,
$$ 
which contradicts the optimality of $(\psi(.),H(.))$, proving Theorem 
\ref{t-main-q2}. \quadp\\\\
{\bf Proof of Theorem \ref{t-main-q1}} In formula \r{kkk1}, the phases 
$\bar \beta^\al_{j,k}$ are arbitrary, so for each interval they could be 
chosen 
to be:\ffoot{magari in questa sezione e' meglio che $\bar H\to\bar H$}
\bqn
\bar\beta_{j,k}^\al=\phi_{j,k}:=arg(\psi_j(0))-arg(\psi_k(0))\in[-\pi,\pi]
\eqnl{PETRASSI}
where $arg(\psi_j(0))$ is intended
to be an arbitrary number if $\psi_j(0)=0$. This proves that if 
$(\psi(.),H(.))$ is a solution to the minimization problem, then there 
exists another solution $(\bar\psi(.),\bar H(.))$ that correspond to 
control $\bar H$ of the form 
\r{rr0},
\r{rr1},
\r{rr2} on 
 $\cup_\al I_\al.$ It remains to consider the 
set $Bad_{j,k}=]0,T[\setminus \cup_\al I_\al $, already defined in formula 
(\ref{bad1}),
\ffoot{i punti estremi 
hanno misura nulla} 
where $\psi_j$ or $\psi_k$ are zero. The following 
Lemma assures that on $Bad_{j,k}$, we can put the control at zero, without 
changing $\bar\psi$ and without increasing the cost.
\bl
Let $(\psi(.),H(.))$ be an admissible pair and define a new control 
$\hat H(.)$ by:
$$
\hat H_{j,k}(t):=\left\{\ba{l}
H_{j,k}(t)\mbox{ ~~for all $t$ such that~~ }\psi_j(t)\psi_k(t)\neq0\\
0\mbox{ ~~otherwise.}
\ea\right.
$$
Then $(\psi(.),{\hat H}(.))$ is also an admissible pair  and its cost is
not bigger than the cost of $(\psi(.),H(.))$.
\llabel{lemma-1}
\el
\proof
Let us first prove that $\hat H(.)$ is a measurable 
function.
Let $N_j$ be the set of times where $\psi_j$ is not 0. Being 
$\psi_j(.)$ measurable it follows that the function
$\un_{j}$ defined by 
$\un_{j}(t)=1$ if $t\in N_j$,
$\un_{j}(t)=0$ if $t\notin N_j$,
is measurable.
This implies that 
$\hat{H}_{j,k}(t)=\un_{j}(t)\un_{k}(t)H_{j,k}(t)$ is measurable and 
$\hat{H}(t)$ 
as well. Let us now recall a classical:\ffoot{chiarire il passaggio dal 
fact alle 
robe dopo}\\\\
{\bf Fact.} Let $g(.):\R\to\R$ be an absolutely continuous function. Then 
the 
set of points $x$ where $g(x)=0$ and $g'(x)\neq0$ has zero measure.\\\\
From this fact, we have a.e.
that $\sum_k H_{j,k}(t)\psi_k(t)=\sum_k \hat H_{j,k}(t)\psi_k(t)$. Indeed, 
when $\psi_k(t)=0$ then $H_{j,k}(t)\psi_k(t)=\hat 
H_{j,k}(t)\psi_k(t)=0$ and for a.e. t such that $\psi_j(t)=0$ we have
$\dot\psi_j(t)=0$ and so $\sum_k H_{j,k}(t)\psi_k(t)=0=\sum_k \hat 
H_{j,k}(t)\psi_k(t)$. Hence
$H(t)\psi(t)=\hat H(t)\psi(t)$ for a. e. $t$ in 
$[0,T]$. It follows that if $\dot\psi(t)=H(t)\psi(t)$ a.e., 
then this 
relation holds also with  $\hat H(t)$ in the place of $H(t)$,
that is $(\psi(.),\hat H(.))$ is an admissible pair.
From the fact that $|\hat H_{j,k}|\leq|H_{j,k}|$, it follows that 
$(\psi(.),{\hat H}(.))$ has a cost 
not bigger than $(\psi(.),H(.))$. The Lemma is proved.
\quadv\\\\
The proof of Theorem \ref{t-main-q1} is now complete. \quadp\\\\
\brem
In formula \r{PETRASSI}, $\bar\beta_{j,k}^\al$ has been chosen in such a 
way to be compatible with the initial condition:
\bqn
\bar\psi(0)=\psi(0).\eqnl{BUSSOTTI}
\erem
\subsection{Transversality}\llabel{s-transversality}
In this Section we give a more geometric interpretation to the proofs of 
Theorems 
\ref{t-main-q1} and
\ref{t-main-q2}.
We are going to show that the vector field 
associated to the control $v_{j,k}^\al$ defined in formula \r{eq-uv} is 
always 
tangent to a submanifold of $S^{2n-1}$ whose points are reached with the 
same cost.  As a consequence the transversality and maximum conditions 
of PMP say  
that the control $v_{j,k}^\al$ can (resp. must)  be set to zero for 
a nondecreasing (resp strictly increasing) cost.
\bl\llabel{lemma2}
Given $\al_1,...,\al_n\in[-\pi,\pi]$, let us define the map: 
$$Rot_{\alpha}:\left\{\ba{rcl}
S^{2n-1} &\to& S^{2n-1}, \\
(\psi_1,...,\psi_n)&\mapsto& (\e^{\ti\alpha_1}\psi_1,...,
\e^{\ti\alpha_n}\psi_n).
\ea\right.
$$ 
If $\psi(.)=(\psi_1(.),\hdots,\psi_n(.))$ defined on $[0,T]$ is an 
admissible curve  then $Rot_{\alpha}(\psi(.))$ is also an 
admissible curve and has
the same cost as $\psi(.)$.
\el
\proof  
It is just a matter of computation. Let us denote by
$H(.)$ the control matrix associated to the admissible curve $\psi(.)$. 
The fact that 
$\psi(.)$ is an admissible curve, associated to the controls $H(.)$, 
writes 
$
\dot{\psi}_k=\sum_j H_{j,k}\psi_j,$ $\forall k\in\{1,\hdots,n\}.
$
By multiplying by $\e^{\ti \alpha_k}$, we find
$
\dot{\psi}_k\e^{\ti \al_k}=\sum_j
H_{j,k}\e^{\ti(\al_k-\al_j)}\psi_j\e^{\ti \al_j}.
$
Hence the curve $Rot_{\alpha}(\psi)$ is an admissible
curve corresponding to  controls 
$\widetilde{H}_{j,k}=H_{j,k}\e^{\ti(\al_k-\al_j)}$.
And since the cost function does not depend on the phase of the
controls, the two curves have the same cost.
\quadp\\\\
This means that $Rot_{\alpha}$
is an ``isometry'' for the cost defined by $f^0$,  for any 
$(\alpha_1,\hdots,\alpha_n)$. A trivial
consequence of this lemma is:
\bc
For any two points $\psi^1$ and $\psi^2$ in $S^{2n-1}$, 
all the points of the set:
$$
{\cal T}_{\psi^2}=\{(\psi^2_1 \e^{\ti
\alpha_1},\hdots, \psi^2_n\e^{\ti \alpha_n})\;
|\;(\alpha_1,\hdots,\alpha_n)\in\R^n \}
$$
are reached with the same cost from the set:
$$
{\cal T}_{\psi^1}=\{(\psi^1_1 \e^{\ti
\alpha_1},\hdots, \psi^1_n\e^{\ti \alpha_n})\;
|\;(\alpha_1,\hdots,\alpha_n)\in\R^n \}.
$$
\ec
As a consequence if  $\psi(.):[0,T]\to S^{2n-1}$ is a minimizing 
trajectory  between the two sets 
${\cal T}_{\psi^1}$
${\cal T}_{\psi^2}$, then the transversality condition of PMP (see {\bf 
v)}) is: 
\bqn
<P(t),T {\cal T}_{\psi(t)}>=0.
\eqnl{eq-t1} 
Now let $F_{j,k}^\al(\psi)$ and $G_{j,k}^\al(\psi)$ defined in 
$I_{j,k}^\al$ be the 
two vector fields associated with 
the 
controls $u_{j,k}^\al$ and $v_{j,k}^\al$ defined in formula  \r{eq-uv}:
$$
\dot\psi_j=\sum_{k}\big(u_{j,k}^\al 
F_{j,k}^\al(\psi)+v_{j,k}^\al G_{j,k}^\al
(\psi)\big)
$$
One can easily check that on $I_\al$:
$$
\ba{l}
{F}_{j,k}^\al(\psi)=\e^{\ti 
\beta_{j,k}^\al}\psi_k\parti{\psi_j}-\e^{-\ti 
\beta_{j,k}^\al}\psi_j\parti{\psi_k},\\
{G}_{j,k}^\al(\psi)=
i (\e^{\ti \beta_{j,k}^\al}\psi_k\parti{\psi_j}+\e^{-\ti 
\beta_{j,k}^\al}\psi_j\parti{\psi_k}).
\ea
$$
And with a simple computation one can see that 
the vector $G_{j,k}^\al(\psi(t))$ is tangent to the set  ${\cal 
T}_{\psi(t)}$: 
\bqn
G_{j,k}^\al(\psi(t))\in T {\cal T}_{\psi(t)},~~~\forall t\in I_\al.
\eqnl{eq-t2}
From \r{eq-t1} and  \r{eq-t2}, one get that 
$<P(t),G_{j,k}^\al(\psi(t))>=0$.

close to any time t of the domain of a given  minimizer,
there exist an interval of time where the minimizer is not strictly
abnormal

there are not strict abnormal extremals. 
Then the maximality condition {\bf iii)} of PMP implies (here 
$f=u_{j,k}^\al F_{j,k}^\al(\psi)+v_{j,k}^\al G_{j,k}^\al(\psi)$):\ffoot{la 
notazione 
e' un po' infelice}
$$\sum_{j,k}
u_{j,k}^\al(t)<P(t), F_{j,k}^\al(\psi(t))>+p_0 
f^0(|u_{j,k}^\al(t)+iv_{j,k}^\al(t)|)=\max_{\bar u_{j,k},\bar v_{j,k}}
\sum_{j,k}\bar u_{j,k}<P(t), F_{j,k}^\al(\psi(t))>+p_0
f^0(|\bar u_{j,k}+i\bar v_{j,k}|)
$$
Now since $p_0<0$ (there are no strictly abnormal extremal, as it will be 
proved in Section \ref{Q3}), we get that
(we stress the fact 
that $v_{j,k}^\al$ is not useful to reach the final target 
since the source and the target are defined by
conditions on the moduli and $G^\al_{j,k}$ is responsible  only of the
evolution of the phases, see eq \r{eq-arg}):
\bi
\i for a non decreasing cost  this condition is  
realized if 
$v^\al_{j,k}=0$ a.e. (plus conditions on $u_{j,k}^\al$);
\i  for a strictly increasing  cost  this condition can be
realized only if
$v^\al_{j,k}=0$ a.e.
\ei
\subsection{Eigenstates}
\llabel{s-pure-states}
Let us recall that an eigenstate is a state  for
which one of the coordinates has norm 1 and the others are 0.

In this Section we show that it is possible 
to join every couple of eigenstates by a minimizing trajectory that is in 
resonance (question {\bf Q1'}). More in general we prove that the answer 
to 
question  {\bf Q1'} is yes for 
any couple of initial and final states $\psi^1=(\psi^1_1,...,\psi^1_n)$ and
$\psi^2=(\psi^2_1,...,\psi^2_n)$ such that 
$\psi^1_j\psi^2_j=0$ for any $j$, that in particular is true for  
eigenstates.

Let $\bar\psi(.)$ be a minimizing trajectory in resonance between the 
points $(|\psi^1_1|,...,|\psi^1_n|)$ and 
$(|\psi^2_1|,...,|\psi^2_n|)$.
Define $\theta^1_j$ and $\theta^2_j$ as the arguments
 of $\psi^1_j$ and $\psi^2_j$, putting them to 0 when the
corresponding coordinate is 0. We define $\alpha_j$ to be equal to 
$\theta^1_j$ if $\psi^1_j\neq 0$, or to $\theta^2_j$ if $\psi^2_j\neq 0$
and 0 if both $\psi^1_j$ and $\psi^2_j$ are 0. Finally let 
$\al:=(\al_1,...,\al_n)$. The curve 
$$\psi(.):=Rot_{\alpha} (\bar\psi(.)),$$
is a resonant minimizer for the problem with initial and final condition 
$\psi^1$ and $\psi^2$. 

\subsection{An Example  of a Non-Weakly-Resonant Minimizer}
\llabel{examples}
In this Section, in the case of a non strictly 
increasing cost, we show an example of an optimal couple 
trajectory-control $(\psi(.),H(.))$, 
joining a source and a target defined by conditions on the moduli,
that is neither resonant nor weakly-resonant.
Consider a time minimization problem for a 4-level system in the form 
(\ref{ham-1}) (i.e. with controls on the 
lower and upper diagonal), for the isotropic case. We have  
$f^0(H)=\max\{|H_{1,2}|,|H_{2,3}|,|H_{3,4}|  \}$, that  
 is not a strictly increasing function but just a not
decreasing one. 

It is easy to see that 
the following curve: $t\mapsto(\cos(t),\sin(t),0,0)$ for 
$t\in[0,\frac{\pi}{2}]$ is a minimizer between the eigenstates 
$|\psi_1|=1$ and 
$|\psi_2|=1$.
It can be obtained 
by different control functions. For instance by $t\mapsto (-1,0,0)$,
or  by $t\mapsto (-1,0,U_3)$ where
$$
\begin{array}{ll}
U_3(t)=1 &\mbox{ for } t\in [0,\frac{\pi}{8}],\\
U_3(t)=-1& \mbox{ for } t\in [\frac{\pi}{8},\frac{\pi}{4}],\\
U_3(t)=i &\mbox{ for } t\in [\frac{\pi}{4},\frac{3\pi}{8}],\\
U_3(t)=-i& \mbox{ for } t\in [\frac{3\pi}{8},\frac{\pi}{2}].
\end{array}
$$
The second one is clearly not weakly-resonant.

The following is still an open question:
Under the hypothesis (H1'), (H2) and  (H3), does there exist an example of 
non-decreasing cost function such that 
there exists a trajectory $(\psi(.),H(.))$, solution of {\bf 
(P')}, being non-weakly-resonant, such that there is no weakly-resonant 
control 
$\bar H$ with $(\psi(.),\bar H(.))$ solution of {\bf (P')} ?
\section{Strictly Abnormal Minimizers for the real problem}
\llabel{Q3}

{\hhh In this Section we prove Theorem \ref{t-main-q3} 
(i.e. we prove 
that close to any time t of the domain of a given  minimizer,
there exist an interval $[t_1,t_2]$ where the minimizer is not strictly
abnormal). We are able to prove the result only in such a interval, 
since on that interval we can make a suitable partition of indexes, see 
Lemma \ref{l-permut} (how to extend this result to the whole domain 
of the minimizer is still an open question, cf. end of Section 
\ref{s-main-results}). Then the difficulty of the proof is coming from the 
fact 
that the dynamics has singularities each time a coordinate is zero.} 
For instance, in the most important example (see formula (\ref{mie})), 
when 
$\psi_1=\psi_2=0$ then the control $V_{1,2}$ has no effect on the 
dynamics, 
i.e. the corresponding vector field is zero when $\psi_1=\psi_2=0$.
Hence, in order to prove the theorem, we 
show (subsections \ref{Q31} and \ref{Q32}) that every
minimizer $\wpsi$ of the problem {\bf (RP')}
is also a solution of an auxiliary optimal problem {\bf (RP'')}, living on 
a submanifold $\J$ of $S^{n-1}$, which has no singularities. Then 
in subsection \ref{Q33}  
we prove 
that this new problem has 
no abnormal extremals (which proves that the 
minimizer $\wpsi$ is a normal extremal as solution of {\bf (RP'')}). 
Finally, we come back to the 
original problem {\bf (RP')} proving that the minimizer has a normal lift 
in $T^*S^{n-1}$ (subsection \ref{Q34}).

\medskip
{\hhh In the following, let $(\wpsi(.),\wH(.))$
be a couple 
trajectory-control 
that
is a minimizer for  {\bf (RP')}.}

\subsection{Permutation of indexes}
\llabel{Q31}

In this subsection, we construct a submanifold $\J$ on which we will 
restrict the optimal problem {\bf (RP')}.

\bl
\llabel{l-permut}
In every \neigh of every time $t$ of the domain of the minimizer, 
there exists a sub-interval 
of time, denoted by $[t_1,t_2]$ (possibly non containing $t$), such 
that there exists a partition $I\cup J$ of $\{1,...,n\}$ satisfying:
\bi
\i if $j\in I$ then $\psi_j(t)=0$ $\forall t\in [t_1,t_2]$,
\i if $j\in J$ then $\psi_j(t)\neq 0$ $\forall t\in [t_1,t_2]$.
\ei
\el
This is just a consequence of the continuity of $\psi(.)$.
The proof is left to the reader. In the following we always restrict to 
the interval $[t_1,t_2]$.

\bdeff
We say that two indexes $j$ and $k$ of $J$ are connected (denoted by 
$j\sim k$) if $j=k$ or there
exists a sequence $j_1,...,j_s$ of indexes of $J$ such that $j=j_1$,
$k=j_s$ and $\forall r<s$ the coefficient $H_{j_r,j_{r+1}}$ is a control. 
\edeff

\brem
This definition of connectedness is exactly the same as in 
subsection \ref{s-graphs} but for the sub-graph defined by $J$. 
Until the end of this Section, "connected" refers to this last definition.
\erem

We immediately get:

\bl
In $J$, the relation "$\sim$" is an equivalence
relation.  
\el
\bdeff
We denote $K_1,...,K_r$ the equivalence classes defined by "$\sim$"
and $m_1,...,m_r$ their cardinalities. We also denote
$M_0=0$ and $M_\ell=\sum_{k\leq\ell} m_k$. 
\edeff
Now, let us make a permutation on the indexes
that orders the sets $K_1,...,K_r,I$ in such a way that: 
$$
\forall \ell\leq r \;\;\;\; K_\ell=\left\{ M_{\ell-1}+1;...;M_{\ell}\right\}, 
$$
and 
$$
I=\left\{ M_r+1;...;n\right\}.
$$

\bl\llabel{lem5}
After the permutation, we have: 
\bd
\i[a)] for all $j\geq M_r+1$ ($j\in I$), $\psi_j\equiv0$ on
$[t_1,t_2]$. 
\i[b)] for all $\ell\leq r$, the map $\displaystyle t\mapsto \sum_{j\in 
K_\ell} |\psi_j(t)|^2=\sum_{j=M_{\ell-1}+1}^{M_\ell} |\psi_j(t)|^2$ 
is constant on $[t_1,t_2]$. 
\ed
\el
\proof 
The point {\bf a)} is just a consequence of the definition of $I$. In 
order to 
prove the point {\bf b)}, let us consider $\ell\leq r$. For any $j$ in 
$K_\ell$, 
we have:
$$
\dot{\wpsi}_j(t)=\sum_{k\leq n} \wH_{j,k}(t)\wpsi_k(t).
$$
But $\wH_{j,k}\equiv 0$ if $k\notin K_\ell\cup I$ and $\wpsi_k\equiv 0$ if 
$k\in I$. Hence for $t$ in $[t_1,t_2]$:
$$
\dot{\wpsi}_j(t)=\sum_{k\in K_\ell} \wH_{j,k}(t)\wpsi_k(t).
$$ 
Now, the matrix $\wH^\ell(t)$, whose coefficients are the $\wH_{j,k}(t)$ 
with 
$j,k \in 
K_\ell$, belongs to $so(m_\ell)$. Hence the vector $\wpsi^\ell$,
whose coefficients are 
the $\wpsi_j(t)$ with $j\in K_\ell$, 
satisfies $\dot{\wpsi}^\ell=\wH^\ell \wpsi^\ell$ and then
has constant norm, i.e. $\sum_{j\in 
K_\ell} |\wpsi_j (t)|^2$ is constant.
\quadp\\\\
From the previous Lemma \ref{lem5}, it follows that for $t\in[t_1,t_2]$, 
$\psi(t)$ belongs to the set:
$$
\J=S^{m_1-1}(C_1)\times...\times S^{m_r-1}(C_r)
\times\prod_{j\in I} \{\psi_j(t_1)\},
$$
where $C_\ell=\sqrt{\sum_{j\in K_\ell}|\psi_j(t_1)|^2}$.

\brem
In the following, we are going to restrict {\bf (RP')} to $\J$. Notice 
that the dimension of $\J$ is $\sum_{\ell\leq r} (m_\ell -1)\leq n-1$,
while the original problem {\bf (RP')} lives on $S^{n-1}$
that has dimension $n-1$.
\erem

\subsection{Restriction of the Control System to $\J$}
\llabel{Q32}
In this Section, we show that $\wpsi$ is also a solution of an auxiliary 
optimal problem {\bf (RP'')} in $\J$.

\bdeff
We call {\bf (RP'')} the optimal problem {\bf (RP')} in which we restrict 
the dynamics by adding a new condition on the matrix of controls: 
$H_{j,k}(t)$ is set to 0 if $j$ or $k$ is in $I$. 
\edeff

\brem
Notice that $\J$ is preserved by the dynamics of {\bf (RP'')}.
\erem

\bl
The curve $\wpsi(.)$ is a minimizer for {\bf (RP'')}.
\el
\proof
As proved in Lemma \ref{lemma-1},
$\wpsi$ does not change if we set to zero the controls 
$\wH_{j,k}$ such that one index belongs to $I$. Hence $\wpsi$ is an 
admissible curve for {\bf (RP'')}.

Now, since the cost function is the same for both {\bf (RP')}
and {\bf (RP'')}, it follows that $\wpsi$ is a minimizer for {\bf 
(RP'')}.

\subsection{The Minimizer $\wpsi(.)$ is not an Abnormal Extremal for 
{\bf 
(RP'')}}
\llabel{Q33}

Let us denote $F_{j,k}(\psi)$ the vector field associated with the control 
$H_{j,k}$: $\dot\psi=\sum_{j,k}H_{j,k}F_{j,k}(\psi)$. 
We denote by $\Delta_\ell$ the 
distribution generated by the $F_{j,k}(\psi)$ with $j,k\in K_\ell$ and 
$\Delta=\oplus_\ell \Delta_\ell$. We have that $\Delta(q)\subset 
T_q\J$ for $q\in\J$.

Using $F_{j,k}(\psi)$, the dynamics of {\bf (RP'')} reads:
$$
\dot{\psi}=\sum_\ell\sum_{j,k\in K_\ell}H_{j,k}F_{j,k}(\psi),
$$

\bl
The distribution $\Delta$ is the whole tangent space of $\J$: $\Delta=T\J$.
\el
\proof
Let us first prove that, for every $\ell$, 
$\Delta_\ell=TS^{m_\ell -1}(C_\ell)$.

\noindent\underline{First step:} it exists a set $L_\ell$ of couples of 
indexes of $K_\ell$, with cardinality $m_\ell -1$, that keeps $K_\ell$ 
connected in the following sense: for any couple of indexes $j,k$ in 
$K_\ell$, there is a sequence 
$j_1,...,j_s$ in $K_\ell$ such that $j_1=j$, $j_s=k$, $\{ j_p,j_{p+1}\}\in 
L_l$ and $H_{ j_p,j_{p+1}}$ is a control.

The proof can be done by induction on the cardinality of $K_\ell$. If it has 
cardinality one, $L_\ell$ is empty. If it has cardinality two, the proof is 
trivial. If the cardinality of $K_\ell$ is bigger than 
two, we 
can choose an index $k\in K_\ell$ such that 
$\widetilde{K}_\ell:=K_\ell-\{k\}$ is still connected 
(this is a standard fact of graph theory). Now, $\widetilde{K}_\ell$ 
is connected and has cardinality $m_\ell-1$. Hence, 
by induction hypothesis, there exists an 
$\widetilde{L}_\ell$ that keeps $\widetilde{K}_\ell$ connected and has 
cardinality $m_\ell-2$. Now, $k$ is connected to an index $j$ of 
$\widetilde{K}_\ell$, hence the set 
$L_\ell=\widetilde{L}_\ell\cup\{\{j,k\}\}$ keeps $K_\ell$ connected and 
has the required cardinality.

\noindent\underline{Second step:} the family of $F_{j,k}(\psi)$ 
($\{j,k\}\in 
L_\ell$) is linearly independent ($L_\ell$ was constructed for this 
purpose).

The proof can be done by induction : if $m_\ell=1$, $L_\ell$ is empty. If 
$m_\ell>1$, then one can choose a index $k$ appearing only once in 
$L_\ell$. Let call $j$ the index such that $\{j,k\}\in L_\ell$. The vector 
$F_{j,k}(\psi)$ is linearly independent of the rest of the family : it is 
the 
only one vector that has a non zero $k$-coordinate.

Now $L_\ell-\{\{j,k\}\}$ has cardinality $m_\ell-2$ and keeps connected 
$K_\ell-\{k\}$ 
of cardinality $m_\ell-1$. Hence we can apply induction.

\noindent\underline{Last step:} From the previous step, it follows that
$\Delta_\ell$ has the same cardinality $m_\ell-1$ as the sphere
$S^{m_\ell-1}(C_\ell)$, hence we get $\Delta_\ell=TS^{m_\ell -1}(C_\ell)$.

Thus we have $\Delta=\oplus_\ell\Delta_\ell=\oplus_\ell TS^{m_\ell 
-1}(C_\ell)=T\J$.\quadp\\\\
Since the distribution $\Delta$ is the whole tangent space of 
$\J$, it is a standard fact that there are no abnormal extremals 
for the problem {\bf (RP'')}.

\subsection{End of the proof}
\llabel{Q34}
Now, we are going to prove that the curve $\wpsi(.)$ admits a normal lift 
on $S^{n-1}$ i.e. it is not a strictly abnormal extremal for 
the problem {\bf (RP')}.

Let us denote $\Theta_\ell$ a local coordinate system on 
$S^{m_\ell-1}(C_\ell)$, and $\Theta=(\Theta_1,...,\Theta_r)$. Then: 
$$(\Theta_1,...,\Theta_r,C_1,...,C_r,\psi_{M_{r+1}},...,\psi_n)$$ 
is a local 
coordinate system on $\R^n$. We denote by $P_{\Theta_\ell}, P_{C_\ell}$ 
and $P_{\psi_j}$ the dual coordinates in $T^*\R^n$, 
$$
P=\sum_{\ell\leq r}P_{\Theta_\ell}d\Theta_\ell + \sum_{\ell\leq 
r}P_{C_\ell}dC_\ell +\sum_{i\in I} P_{\psi_i}d\psi_i,
$$
and
$$
P_\J=\sum_{\ell\leq r}P_{\Theta_\ell}d\Theta_\ell
$$
its restriction to $\J$.

\medskip

\brem\label{remarque}
Notice that the vector fields 
$F_{j,k}$ with $j,k$ in $K_\ell$, depend only on the $\Theta_\ell$ 
coordinates and that they annihilate $dC_s$ ($s\leq r$) and $d\psi_i$ 
($i\in I$). This fact will be essential to conclude the proof.
\erem

The PMP-Hamiltonian associated with {\bf (RP'')} is:
$$
{\HHH}_\J(\Theta,H,P_\J,p_0)=
P_\J(\sum_{j,k\in J}H_{j,k}F_{j,k}(\Theta))+p_0f^0(H),
$$
where $f^0$ is the cost function. Since $\wpsi$ is a normal extremal for 
{\bf (RP'')}, there is a lift 
$(\wpsi,\widetilde{H},\widetilde{P}_\J,\widetilde{p}_0)$, 
with 
$\widetilde{p}_0\neq 0$, that satisfies:
$$
\begin{array}{ccc}
\dot{\widetilde{\Theta}}_\ell & = & \frac{\partial {\HHH}_\J}{\partial 
P_{\Theta_\ell}}
(\widetilde{\Theta},\widetilde{H},\widetilde{P}_\J,\widetilde{p}_0)\\
\dot{\widetilde{P}}_{\Theta_\ell} & = & -\frac{\partial 
{\HHH}_\J}{\partial \Theta_\ell}
(\widetilde{\Theta},\widetilde{H},\widetilde{P}_\J,\widetilde{p}_0)
\end{array}
$$
and the maximality condition:
$${\HHH}_\J(\wpsi(t),\widetilde{H}(t),\widetilde{P}_\J(t),\widetilde{p}_0)
=\max_H\{{\HHH}_\J(\wpsi(t),H,\widetilde{P}_\J(t),\widetilde{p}_0)\},$$
where the maximum is taken over the set of controls.

Now, the PMP-Hamiltonian associated with {\bf (RP')} is:
$$
{\HHH}(\psi,H,P,p_0)=
P(\sum_{j,k\leq n}H_{j,k}F_{j,k})+p_0f^0(H).
$$
To conclude the proof of Theorem \ref{t-main-q3} we need the 
following:\\\\
\noindent{\bf Claim}
{\it Let us denote:  
$\widetilde{P}=\widetilde{P}_\J+\sum_{\ell\leq r} 0\; dC_\ell +\sum_{i\in 
I} 0\;d\psi_i$.
Then $(\wpsi,\widetilde{H},\widetilde{P},\widetilde{p}_0)$
is a normal lift of $\wpsi$ satisfying the PMP for {\bf (RP')}}.\\ \\
\proof
Let us first remark that $\widetilde{P}$ satisfies 
$\widetilde{P}_{C_\ell}=\widetilde{P}_{\psi_i}=0$ for every $\ell\leq r$ 
and $i\in I$.

In order to prove the claim, we have to prove that the lift satisfies 
the Hamiltonian equations: 
\begin{eqnarray}
\dot\wpsi & = & \frac{\partial \HHH}{\partial P}
(\wpsi,\wH,\widetilde{P},\widetilde{p}_0) \label{jacob1}\\ 
\dot{\widetilde{P}} & = & -\frac{\partial \HHH}{\partial \psi} 
(\wpsi,\wH,\widetilde{P},\widetilde{p}_0)\label{jacob2}
\end{eqnarray}
and the maximality condition: 
\begin{eqnarray}
{\HHH}(\wpsi(t),\widetilde{H}(t),\widetilde{P}(t),\widetilde{p}_0)
=\max_H\{{\HHH}(\wpsi(t),H,\widetilde{P}(t),\widetilde{p}_0)\}. 
\label{maxcond}
\end{eqnarray}

The equation (\ref{jacob1}) is always trivially satisfied (it is the 
dynamics).

Let us prove that (\ref{jacob2}) holds. Firstly, we have that:
$$
\dot{\widetilde{P}}_{\Theta_\ell}=-\frac{\partial \HHH_\J}{\partial 
\Theta_\ell}(\widetilde{\Theta},\widetilde{H},\widetilde{P}_\J,
\widetilde{p}_0)=-\frac{\partial \HHH}{\partial \Theta_\ell}
(\wpsi,\widetilde{H},\widetilde{P},\widetilde{p}_0).
$$
Secondly, 
$$
\frac{\partial \HHH}{\partial C_\ell}
(\wpsi,\widetilde{H},\widetilde{P},\widetilde{p}_0)
=\widetilde{P}(\sum_{j,k\leq n} \widetilde{H}_{j,k} 
\frac{\partial}{\partial C_\ell} (F_{j,k}))
$$
but $\widetilde{H}_{j,k}=0$ if $i$ or $k$ is in $I$ and, because of Remark
\ref{remarque}, $\frac{\partial}{\partial C_\ell} (F_{j,k})=0$ when $j$ 
and $k$ are in $J$. Hence $\dot{\widetilde{P}}_{C_\ell}=0=-\frac{\partial 
\HHH}{\partial 
C_\ell}(\wpsi,\widetilde{H},\widetilde{P},\widetilde{p}_0)$.
The proof is the same for $\widetilde{P}_{\psi_i}$ for $i\in I$. Hence
(\ref{jacob2}) is satisfied.

Let us now prove (\ref{maxcond}). We have that, if $j$, or $k$, is in $I$ 
then $F_{j,k}(\wpsi)$ 
is in $\mbox{span}\{\frac{\partial}{\partial \psi_i}\; ;\; i\in I \}$. 
Indeed, $F_{j,k}(\wpsi)=\wpsi_k\frac{\partial}{\partial 
\psi_j}-\wpsi_j\frac{\partial}{\partial \psi_k}$, and then if $j\in I$, 
then $\wpsi_j=0$ and $F_{j,k}(\wpsi)=\wpsi_k\frac{\partial}{\partial
\psi_j}$. Hence, because $\widetilde{P}_{\psi}=0$, we have that
$\widetilde{P}(F_{j,k})=0$ if $j$ or $k$ is in $I$ and we have 
that for any control $H$:
$$
\begin{array}{ccl}
{\HHH}(\wpsi,H,\widetilde{P},\widetilde{p}_0)& = & \displaystyle
\widetilde{P}(\sum_{j,k\leq n}
H_{j,k}F_{j,k}(\wpsi))+\widetilde{p}_0f^0(H)\\ 
& = & \displaystyle
\widetilde{P}(\sum_{j,k\in 
J}H_{j,k}F_{j,k}(\wpsi))+\widetilde{p}_0f^0(H)\\
 & = & \displaystyle \widetilde{P}_\J(\sum_{j,k\in 
J}H_{j,k}F_{j,k}(\wpsi))+\widetilde{p}_0f^0(H)\\
& = & {\HHH}_\J(\wpsi,H,\widetilde{P}_\J,\widetilde{p}_0)
\end{array}
$$
Now the fact that 
$(\wpsi,\widetilde{H},\widetilde{P}_\J,\widetilde{p}_0)$
satisfies the maximality condition for {\bf (RP'')} allows to conclude 
that $(\wpsi,\widetilde{H},\widetilde{P},\widetilde{p}_0)$ satisfies 
the maximality condition for {\bf (RP')}.
\quadp

\medskip
\noi
{\bf AKNOWLEDGMENTS} \\
{\hhh The authors are grateful to Andrei Agrachev and 
Jean-Paul Gauthier for many helpful discussions. The authors would like 
also to thank the anonymous referee for some crucial remarks.}


\fine

%% file: figura1.pstex_t
\begin{picture}(0,0)%
\includegraphics{figura1.pstex}%
\end{picture}%
\setlength{\unitlength}{2763sp}%
\begingroup\makeatletter\ifx\SetFigFont\undefined%
\gdef\SetFigFont#1#2#3#4#5{%
  \reset@font\fontsize{#1}{#2pt}%
  \fontfamily{#3}\fontseries{#4}\fontshape{#5}%
  \selectfont}%
\fi\endgroup%
\begin{picture}(11079,14394)(724,-14008)
\put(1576,-436){\makebox(0,0)[lb]{\smash{\SetFigFont{11}{13.2}{\rmdefault}{\mddefault}{\updefault}Isotropic Energy: $f^0=|V_{1,2}|^2+|V_{2,3}|^2$}}}
\put(2326,-10561){\makebox(0,0)[lb]{\smash{\SetFigFont{11}{13.2}{\rmdefault}{\mddefault}{\updefault}{\color[rgb]{0,0,0}(Isotropic Case)}%
}}}
\put(7951,-10561){\makebox(0,0)[lb]{\smash{\SetFigFont{11}{13.2}{\rmdefault}{\mddefault}{\updefault}{\color[rgb]{0,0,0}(Non-isotropic Case)}%
}}}
\put(6901,-10111){\makebox(0,0)[lb]{\smash{\SetFigFont{11}{13.2}{\rmdefault}{\mddefault}{\updefault}{\color[rgb]{0,0,0}Time with b. contr: $f^0=\max(\frac{|V_{1,2}|}{\mu_{1,2}},\frac{|V_{2,3}|}{\mu_{2,3}})$}%
}}}
\put(3376,-1111){\makebox(0,0)[lb]{\smash{\SetFigFont{11}{13.2}{\rmdefault}{\mddefault}{\updefault}{\color[rgb]{0,0,0}$V_{2,3}$}%
}}}
\put(9946,-2806){\makebox(0,0)[lb]{\smash{\SetFigFont{11}{13.2}{\rmdefault}{\mddefault}{\updefault}$V_{1,2}$}}}
\put(4426,-2761){\makebox(0,0)[lb]{\smash{\SetFigFont{11}{13.2}{\rmdefault}{\mddefault}{\updefault}$V_{1,2}$}}}
\put(3286,-6166){\makebox(0,0)[lb]{\smash{\SetFigFont{11}{13.2}{\rmdefault}{\mddefault}{\updefault}{\color[rgb]{0,0,0}$V_{2,3}$}%
}}}
\put(4246,-7696){\makebox(0,0)[lb]{\smash{\SetFigFont{11}{13.2}{\rmdefault}{\mddefault}{\updefault}$V_{1,2}$}}}
\put(9481,-7711){\makebox(0,0)[lb]{\smash{\SetFigFont{11}{13.2}{\rmdefault}{\mddefault}{\updefault}$V_{1,2}$}}}
\put(10006,-12691){\makebox(0,0)[lb]{\smash{\SetFigFont{11}{13.2}{\rmdefault}{\mddefault}{\updefault}$V_{1,2}$}}}
\put(4201,-12706){\makebox(0,0)[lb]{\smash{\SetFigFont{11}{13.2}{\rmdefault}{\mddefault}{\updefault}$V_{1,2}$}}}
\put(6751,-436){\makebox(0,0)[lb]{\smash{\SetFigFont{11}{13.2}{\rmdefault}{\mddefault}{\updefault}Non-isotropic Energy: $f^0=\frac{|V_{1,2}|^2}{\mu_{1,2}^2}+\frac{|V_{2,3}|^2}{\mu_{2,3}^2}$}}}
\put(1276,-10111){\makebox(0,0)[lb]{\smash{\SetFigFont{11}{13.2}{\rmdefault}{\mddefault}{\updefault}{\color[rgb]{0,0,0}Time with b. contr: $f^0=\max(|V_{1,2}|,|V_{2,3}|)$}%
}}}
\put(1651,-5311){\makebox(0,0)[lb]{\smash{\SetFigFont{11}{13.2}{\rmdefault}{\mddefault}{\updefault}Isotropic Area: $f^0=|V_{1,2}|+|V_{2,3}|$}}}
\put(9151,-1216){\makebox(0,0)[lb]{\smash{\SetFigFont{11}{13.2}{\rmdefault}{\mddefault}{\updefault}{\color[rgb]{0,0,0}$V_{2,3}$}%
}}}
\put(8761,-6226){\makebox(0,0)[lb]{\smash{\SetFigFont{11}{13.2}{\rmdefault}{\mddefault}{\updefault}{\color[rgb]{0,0,0}$V_{2,3}$}%
}}}
\put(9376,-11101){\makebox(0,0)[lb]{\smash{\SetFigFont{11}{13.2}{\rmdefault}{\mddefault}{\updefault}{\color[rgb]{0,0,0}$V_{2,3}$}%
}}}
\put(3376,-11161){\makebox(0,0)[lb]{\smash{\SetFigFont{11}{13.2}{\rmdefault}{\mddefault}{\updefault}{\color[rgb]{0,0,0}$V_{2,3}$}%
}}}
\put(6376,-5311){\makebox(0,0)[lb]{\smash{\SetFigFont{11}{13.2}{\rmdefault}{\mddefault}{\updefault}Non-isotropic Area: 
$f^0=\frac{|V_{1,2}|}{\mu_{1,2}}+\frac{|V_{2,3}|}{\mu_{2,3}}$}}}
\end{picture}